\documentclass[a4paper,fleqn,usenatbib]{mnras}

\usepackage[T1]{fontenc}
\usepackage{ae,aecompl}

\usepackage{graphicx}	
\usepackage{amsmath}	
\usepackage{amssymb}	

\usepackage[usenames, dvipsnames]{color}
\newcommand{\Amy}{\color{Black} }

\title[Dwarf Planet Densities]{Interpreting the Densities of the Kuiper Belt's Dwarf Planets}

\author[Barr and Schwamb]{
Amy C. Barr$^{1}$\thanks{E-mail: amy@psi.edu (ACB)}
and Megan E. Schwamb$^{2}$
\\
$^{1}$Planetary Science Institute, 1700 E. Fort Lowell, Suite 106, Tucson, AZ 85719, United States\\
$^{2}$Institute for Astronomy and Astrophysics, Academia Sinica; 11F AS/NTU, \\ National Taiwan University, 1 Roosevelt Rd., Sec. 4, Taipei 10617, Taiwan\\
}

\date{Accepted April 29, 2016. Received April 29, 2016; in original form March 13, 2016}

\pubyear{2016}

\begin{document}
\label{firstpage}
\pagerange{\pageref{firstpage}--\pageref{lastpage}}
\maketitle

\begin{abstract}
Kuiper Belt objects with absolute magnitude less than 3 (radius $\gtrsim$500 km), the dwarf planets, have a range of different ice/rock ratios, and are more rock-rich than their smaller counterparts.  Many of these objects have moons, which suggests that collisions may have played a role in modifying their compositions.  We show that the dwarf planets fall into two categories when analysed by their mean densities and satellite-to-primary size ratio.  Systems with large moons, such as Pluto/Charon and Orcus/Vanth, can form in low-velocity grazing collisions in which both bodies retain their compositions.  We propose that these systems retain a primordial composition, with a density of about 1.8 g/cm$^3$.  Triton, thought to be a captured KBO, could have lost enough ice during its early orbital evolution to explain its rock-enrichment relative to the primordial material.  Systems with small moons, Eris, Haumea, and Quaoar, formed from a different type of collision in which icy material, perhaps a few tens of percent of the total colliding mass, is lost.  The fragments would not remain in physical or dynamical proximity to the parent body.  The ice loss process has not yet been demonstrated numerically, which could be due to the paucity of KBO origin simulations, or missing physical processes in the impact models.  If our hypothesis is correct, we predict that large KBOs with small moons should be denser than the primordial material, and that the mean density of Orcus should be close to the primordial value.
\end{abstract}




\section{Introduction}
The Kuiper belt is composed of icy building blocks left over after the era of planet formation beyond the orbit of Neptune. The distant dwarf planets, Kuiper belt objects with absolute magnitude less than 3 (radii $\gtrsim 500$ km), represent the largest products of planetesimal accretion in the outer Solar System, with the vast majority of Kuiper belt objects (KBOs) being much smaller in size. Wide-field surveys \citep{Millis2002, Trujillo2003, Elliot2005, Jones2006, Larsen2007,  BrownTNOBook, Petit2011,Schwamb2010, Sheppard2011, Rabinowitz2012, Brown2015,Bannister2016} have completed the inventory of distant dwarf planets brighter than $\sim$21.5th apparent R magnitude \citep{Schwamb2013, Brown2015}, yielding seven bodies comparable in size to Pluto: Makemake, Haumea, Eris, Sedna, 2007 OR10, Quaoar, and Orcus. These objects are bright enough to be studied with the current suite of large ground-based and space-based telescopes. Observations of their sizes, masses, and compositions can provide a window into the early environment of the outer Solar System and planet formation processes \citep[e.g.,][and references therein]{BrownTNOBook,BrownReview2012}.

%
Adaptive optics observations with ground-based 8-10-m class telescopes and imaging from the Hubble Space Telescope have revealed that the majority of the dwarf planets have satellites, and that satellites are more common in the dwarf planet population than in other size classes \citep{Brown2006}.  A moon's orbit about its primary and knowledge of the primary's size allow an estimate of density.  The densities of the dwarf planets with moons range from $\sim$1.6 g/cm$^3$ to 2.6 g/cm$^3$ \citep{Tholen2008, Brown2010Orcus, Lellouch2010, SicardyNature2011, Braga-RibasQuaoar, LockwoodHaumea, Stern2015}, suggesting that some are more rock-rich than others. The data reveal no clear correlation between size and density \citep{BrownTNOBook}, and a number of oddities, notably, Eris, which is 27\% more massive than Pluto, despite having a nearly identical radius \citep{Brown2007Eris,SicardyNature2011}. 

In the past six years, improvements in the predictions of stellar occultations, and \emph{New Horizons} images of the Pluto system, have provided updated information about the sizes and densities of the dwarf planets.  The radii derived from these methods have uncertainties of a few kilometres, in contrast to the much larger measurement errors from optical and infrared observations.  Images of Pluto from the \emph{New Horizons} flyby in July 2015 have yielded an extremely accurate measurement of Pluto's physical size, separate from its atmosphere \citep{Stern2015}. Before the encounter, the best estimates of the densities of Pluto and Charon were $\bar{\rho}=2.06$ g/cm$^3$ and $\bar{\rho}=1.63$ g/cm$^3$ \citep{Tholen2008}.  Post-encounter densities for the two bodies differ by only 10\% \citep{Stern2015}, and the value of Pluto's density has been lowered to 1.86 $\pm$0.01 g/cm$^3$.  Recent stellar occultation data show that Quaoar, once thought to be essentially pure rock \citep{FraserQuaoar2010, Fraser2013}, is more ice-rich, with $\bar{\rho}=1.99 \pm 0.46$ g/cm$^3$ \citep{Braga-RibasQuaoar}.

Why are the dwarf planets so rock-rich, and why do they exhibit such a large range of densities?  The estimated densities of smaller KBOs, like Salacia (radius $\sim 430$ km, $\bar{\rho} =1.29^{+0.29}_{-0.23}$ g/cm$^3$ \citep{Fornasier2013}), and 2002 UX25 (radius $\sim 325$ km, $\bar{\rho}=0.82 \pm 0.11$ g/cm$^3$ \citep{Brown2002UX25}), indicate low rock fractions, even when porosity is accounted for \citep{Brown2002UX25}.  Thus, simply accreting large rock-rich KBOs from these smaller, rock-poor bodies is not a viable scenario \citep{Brown2002UX25}, and other processes must be at work.

The presence of moons around the dwarf planets points toward planet-scale collisions as a means of changing KBO densities \citep{BrownFamily2007, BrownTNOBook, Brown2002UX25}.  Collisions between large KBOs are thought to be common, occurring every $\sim 100$ to 300 Myr \citep{KenyonBromley2013} throughout the history of the solar system.  The high angular momentum per unit mass of the Pluto/Charon system suggests that the system formed via an impact \citep{Canup2005}.  There is direct evidence that Haumea was modified by a collision: Haumea is the progenitor of a collisional family of ice-rich and neutral-coloured objects identified in the Kuiper belt \citep{BrownFamily2007, Snodgrass2010, Carry2012}, thought to be fragments of the mantle removed from the body.  In its present state, Haumea has a water ice-rich spectrum but a rock-rich composition, and is spinning at nearly break-up speed.  

Here, we show that the dwarf planet systems fall into two categories, when characterised by the bulk density of their primary object and the size ratio between moons and primary.  We hypothesise that the two types of systems formed in two types of collisions.  Low-velocity collisions between undifferentiated primordial dwarf planets make large planet/moon pairs, in which both bodies retain their original compositions, similar to Pluto/Charon.  Higher velocity collisions between fully differentiated dwarf planets could yield rock-enriched primaries with small ice-rich satellites.  

\section{Dwarf Planet System Properties}
 We consider KBOs with a primary or secondary whose effective radius is greater than $\sim$500 km or absolute magnitude (H) less than 3, the magnitude at which the number and sizes of objects begins to deviate from a power law \citep{BrownTNOBook}.  At the time of writing, this consists of five dwarf planets with semi-major axes $>30$ AU, for which we have mass and size measurements: Pluto, Quaoar, Eris, Orcus, and Haumea. Of the satellites of the dwarf planet systems, only Charon, Pluto's largest moon, satisfies these requirements and is included in our sample.  Previous works have compared the icy dwarf planets to  Triton, an irregular satellite of Neptune, thought to have originated in the same region as the largest KBOs and been captured by Neptune \citep{McKinnon1984, GoldreichNeptune, AgnorPluto,Nogueira2011}.  

Table \ref{table:properties} summarises the radii and mean densities of each of these objects.  We calculate the density of the primary object from the system mass and the best measured effective primary radius.  System masses are taken from \citet{Brown2007Eris,RagozzineBrown2009,Brown2010Orcus,Fraser2013} and \citet{Brozovic2015}.  If stellar occultations or \emph{New Horizons} measurements are available, we use those size estimates for the primary body \citep{SicardyNature2011,Braga-RibasQuaoar,Stern2015}.  For Orcus/Vanth, we use estimates of sizes based on modeling Spitzer observations \citep{Brown2010Orcus}.  Otherwise, the most recent Herschel observations for the effective system radius are used \citep{Fornasier2013}.  

\begin{table}
	\centering
	\caption{Radii and mean densities for the dwarf planets plus Neptune's satellite Triton, thought to be a captured dwarf planet.   	
$^{a}$Best-fitting shapes are ellipsoids/spheroids; radius of sphere with equivalent volume.  $^b$Assuming Orcus and Vanth have equal albedo. $^{c}$Assuming the albedo of Vanth is half that of Orcus \citep{Brown2010Orcus}.}
	\label{table:properties}
	\begin{tabular}{lll} 
		\hline
		Body & Radius (km) & $\rho$ (g/cm$^3$) \\
		\hline
		Triton & 1353.4$\pm$0.9 & 2.061  \\
		Eris 				& 1163$\pm$6 & 2.52$\pm$0.05 \\	 
		Haumea$^{a}$ & 620 $\pm{^{34}_{ 29}}$ & 2.6 \\			 
		Orcus$^b$ 			& 430 	        &    1.605$\pm$0.03 	 \\	 
		Orcus	$^c$			& 450		& 1.747$\pm$0.03 \\
		Pluto 				& 1187$\pm$4 & 1.86 $\pm$0.01   \\   
		Charon 			& 606$\pm$3 & 1.702$\pm$0.02 \\ 
		Quaoar$^a$ & 555$\pm 2.5$ & 1.99$\pm$0.46  \\
		\hline
	\end{tabular}
\end{table}

For the Pluto-Charon system, the exact masses and densities are both bodies are known \citep{Brozovic2015}.  We do not yet know the relative sizes of Orcus and its moon Vanth.  The thermal emission from the total system is consistent with a single body of radius 470$\pm35$ km \citep{Brown2010Orcus}.  But estimates of the sizes of each body requires that assuming albedoes for each body.  In Table \ref{table:properties}, we report two values for the radius of Orcus, one assuming Vanth has an albedo equal to that of Orcus \citep{Brown2010Orcus}, which yields a density for both members of the system $\bar{\rho}=1.605 \pm 0.03$ g/cm$^3$.  If Vanth has an albedo half that of Orcus, the radius of Orcus is 430 km, yielding a density for Orcus and Vanth $\bar{\rho}=1.747 \pm 0.03$ g/cm$^3$ \citep{Brown2010Orcus}.  The assumption of unequal albedoes seems consistent with the disparate surface compositions of the objects: Vanth does not share the water absorption feature present in the infrared spectrum of Orcus, indicating that the two bodies have different surface compositions \citep{Brown2000, Brown2010Orcus}.  Both pairs of size estimates for Orcus and Vanth are similar to the sizes derived from Herschel data \citep{Fornasier2013}.

For Eris, Haumea, and Quaoar, we know the system mass; the satellites are so small that assumptions about their physical properties do not affect the estimate of primary mass \citep{Brown2007Eris,RagozzineBrown2009,Fraser2013}.  However, it is worth noting that the primaries and moons in each of these systems have similar colour, in contrast to Orcus and Vanth (e.g., \citealt{BrownTNOBook} and references therein).  The large uncertainty in the density for Quaoar is due to the uncertainty in Quaoar's mass \citep{Fraser2013,Braga-RibasQuaoar}.  At present, there are no independent estimates for size and density for Haumea.  The current estimates for its size and density, which come from combining optical and thermal data \citep{Lellouch2010,LockwoodHaumea}, report only their best-fitting results, and give the same density of 2.6 g/cm$^3$.  Haumea has a spin period of 3.9 days \citep{Lacerda2008}, and has a shape best fit by a tri-axial ellipsoid \citep{Rabinowitz2006,Lellouch2010, LockwoodHaumea}.  We use the effective radius measured from Hershel observations \citep{Fornasier2013}.

\begin{table}
\centering
	\caption{Satellite-to-total system mass ratio, $q$ for each of the dwarf planet systems. $^a$Assuming Orcus and Vanth have equal albedo. $^{b}$Assuming the albedo of Vanth is half that of Orcus \citep{Brown2010Orcus}.}
	\label{table:qvalues}
	\begin{tabular}{lll}
		\hline
		Primary & Satellite & \multicolumn{1}{c}{$q$}\\ 
		\hline
		Eris         & Dysnomia &  0.0253$\pm$0.006\\ 
		Haumea & Hi'iaka &  0.0049$\pm$0.007\\   
 		Orcus$^a$      & Vanth  &  $\sim0.0292$\\
		Orcus$^b$     & Vanth &  $\sim0.0794$\\
		Pluto       & Charon &  0.1086$\pm$0.001\\
		Quaoar   & Weywot &  0.00053$\pm$0.0002\\	
	  	\hline
	\end{tabular}
\end{table}

\begin{figure}
	\includegraphics[width=\columnwidth]{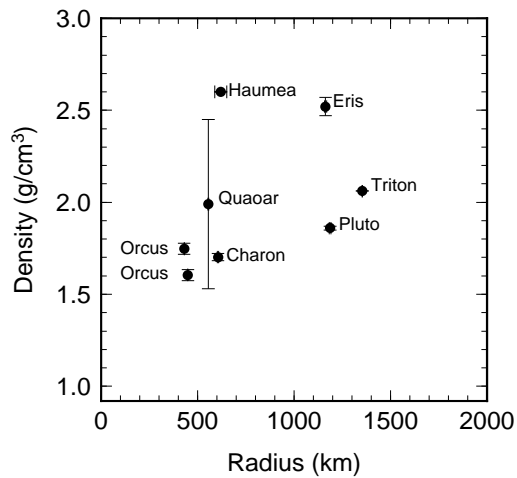}
    \caption{Updated values for the densities and sizes of the dwarf planets, including Neptune's satellite Triton.}
   \label{fig:densities}
\end{figure}

Figure \ref{fig:densities} illustrates the sizes and densities of the dwarf planet primaries.  \citet{BrownTNOBook} notes no statistically significant correlation between size and density; the updated masses and radii determined in the last few years seem to support this conclusion.  We are also interested in the relative sizes of the satellites versus the primary, which we describe as $q=M_{satellites}/(M_{satellites}+M_p)$, the ratio between the combined masses of the satellites and the total system mass, where $M_p$ is the mass of the primary.  

Table \ref{table:qvalues} summarises the $q$ values for each of the systems.  To derive $q$, we assume equal albedos and densities for all of the bodies in the Eris and Quaoar systems.  Thus, our values of $q$ are upper limits; if the small satellites are more ice-rich than their primaries, the true $q$ values for Eris and Quaoar could be smaller.  For Haumea, we use estimates of the masses of each satellite from the multi-body orbital fits of \citet{RagozzineBrown2009}.  For Orcus/Vanth, the assumption of equal albedo is known to be inaccurate.  Following \citep{Brown2010Orcus}, we report two values of $q$, one assuming the bodies have the same albedo, and the other where we assume the albedo of Vanth is a factor of two lower than Orcus, which may be more consistent given the colour difference observed between the two bodies.   

\section{Two Classes of Dwarf Planet Densities}
\begin{figure}
	\includegraphics[width=\columnwidth]{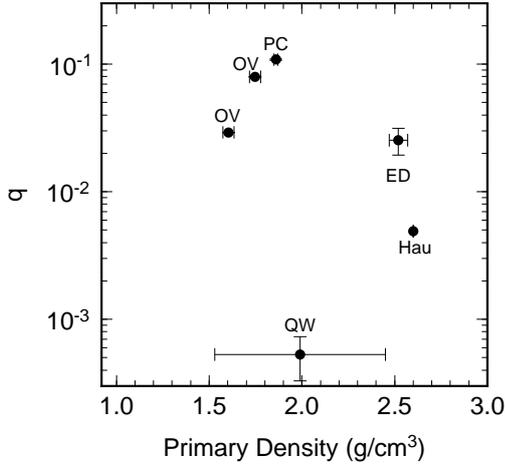}
    \caption{Satellites-to-total mass ratio, $q$ as a function of primary density for the dwarf planet systems Pluto/Charon, Orcus/Vanth, Eris/Dysnomia, Quaoar/Weywot, and Haumea.}
   \label{fig:families}
\end{figure}

We find that the systems fall into two categories when classified by $q$ (see Figure \ref{fig:families}). Excluding Triton, bodies with densities $>$2 g/cm$^3$, Haumea, Quaoar, and Eris, have extremely small satellites, indicated by small $q$ values.  More ice-rich bodies, including Pluto, Charon, and Orcus, are part of systems with large moons.  In a collision, the mass of the final moon and the system's final $q$ value depend on the speed, angle, and impactor-to-total mass ratio, and the differentiation state of the precursor objects (see, e.g., \citealt{Leinhardt2012I} for discussion).  

We hypothesise that the two categories represent two classes of collision, each of which has a different effect on the densities of the final objects.  Systems with large moons originate in low-velocity, grazing collisions between undifferentiated precursors.  These collisions involve little-to-no vaporisation or melting, and so both bodies retain their primordial compositions.  Thus, the primordial composition of Kuiper Belt material can be inferred from the present compositions of members of these systems.  Systems with rock-rich primaries and small moons originate in a different type of collision (see \citealt{BrownTNOBook} and references therein), perhaps a ``graze and merge'' \citep{LeinhardtHaumea}, or another type of collision yet to be identified by numerical simulations.  

\subsection{Lower Density, Large Moons}
\subsubsection{Pluto/Charon}
Successful {\Amy hydrocode simulations of} impact scenarios for the formation of the Pluto system involve an impactor-to-total mass ratio, $\gamma=0.3$ to 0.5, implying that precursor bodies range from 0.3 $M_T$ to $0.7M_T$ \citep{Canup2011}, where the total mass of the Pluto/Charon system, $M_T=1.463 \times 10^{25}$ grams \citep{Stern2015}.  
The successful collisions are gentle, with impact velocities $v_{imp}\approx v_{esc}$, where $v_{esc}=\sqrt{(2GM_T)/(R_{i}+R_{t})}$, where $R_i$ and $R_t$ are the radii of the impactor and target, respectively \citep{Canup2011}; for the Pluto/Charon system, $v_{esc}\sim$ 1 km/s.
The peak shock pressure at the point of impact, $P\sim \rho v_{imp}^2 \sim 1$ GPa \citep{Melosh1989}, is barely high enough to melt ice \citep{StewartAhrens2005, melt_volumes}.  Impact simulations show that the temperature rise in the interiors of both bodies $\Delta T \sim$ tens of K \citep{Canup2005}.  The successful collision is so oblique that the bodies do not undergo significant mixing or modify their original densities.  

%

\citet{Canup2011} shows that the mixed ice/rock composition of Charon is best reproduced by impacts in which one or both objects are undifferentiated {\Amy (see also \citealt{Desch2015})}. 
The most successful cases are those in which 90\% of both precursor bodies are composed of an intimate mixture of ice and rock, with the remaining 10\% of their masses composed of pure water ice, likely in an outer ice shell \citep{Canup2011}.  
If both bodies have undergone complete ice/rock separation, their rock cores merge during the collision, and the relatively small amount of material launched into orbit is too ice-rich to account for Charon's mean density \citep{Canup2005,Canup2011}.  
The small ice-rich satellites of Pluto, Styx, Nix, Kerberos, and Hydra, represent fragments of the pure ice mantles of the precursor bodies launched into orbit by the collision \citep{Canup2011, KenyonBromley2013}.

\subsubsection{Orcus/Vanth}
The Orcus/Vanth system shares several similarities to Pluto/Charon, including its $q$ value and dynamical state.
Vanth orbits Orcus at a distance of $a=8980\pm$20 km \citep{Brown2010Orcus}, $\sim20$ Orcus radii, similar to Charon's orbit, at $\sim15$ Pluto radii.  
Like Pluto and Charon, Orcus and Vanth show different spectral features indicating slightly different surface composition \citep{Brown2010Orcus,Carry2011}.  

Two possible modes of origin have been suggested for the system: a giant impact followed by tidal evolution to a low-eccentricity orbit and capture into a high-inclination orbit followed by damping by the Kozai mechanism \citep{Brown2010Orcus}.  
If the system formed as a result of a giant impact, the masses of Orcus and Vanth are high enough to permit the system to have evolved into at least a synchronous state (in which the orbital period of Vanth is equal to its rotation period) \citep{Brown2010Orcus}.
%

If the system has achieved its final dual synchronous state, the angular momentum of the system about its centre of mass \citep{Canup2005},
\begin{equation}
L_{sys}\approx \frac{q \omega M_{sys}a^2}{(1+q)^2},
\end{equation}
where $M_{sys}$ is the combined masses of Orcus and Vanth, and $a$ is the semi-major axis of Vanth's orbit.  
This is commonly scaled by $L'=(GM^3_{sys}R_{sys})^{1/2}$, where $R_{sys}=[M_{sys}/(4/3 \pi \bar{\rho})]^{1/3}$.  
The normalised system angular momentum, $J=L_{sys}/L'$ \citep{Canup2005}.  
For Orcus/Vanth, $J\approx0.12$ to 0.30.  By comparison, $J\approx 0.28$ for post-encounter system properties for Pluto/Charon, and $J\approx 0.11$ for the Earth/Moon system \citep{Canup2004}.

Thus, it seems plausible that the system formed as a result of a giant impact, similar in character to the Pluto/Charon collision.
With these values of $q$ and $J$, the Orcus/Vanth system falls into the range of impact outcomes reported by \citet{Canup2005} in her initial search for the Pluto/Charon impact.  
Two of the outcomes from \citet{Canup2005} yield the proper $q$ and $J$ for Orcus/Vanth, but were performed for larger $M_T$. 
The candidate collisions occur between undifferentiated objects, with $\gamma=0.3$, $v_{imp} \sim 1.1$ to 1.2$v_{esc}$, which would correspond to about 0.3 to 0.4 km/s for Orcus/Vanth.  
Thus, the system could have formed from a collision between undifferentiated objects ranging in size from 300 to 400 km.  

Regardless of its mode of origin, Orcus and Vanth have likely retained much of their primordial compositions.  
If Vanth were captured, modification of its orbit by tidal friction would cause an average temperature rise of only $\sim 1$ to 10 K for the duration of the orbital evolution \citep{RaggozineThesis}.  
The candidate collisions would result in little-to-no mass loss, melting, or vaporisation, and little heating \citep{Canup2005}, preserving the primordial compositions of both bodies.
This may explain why Orcus and Vanth have vastly different colours, unlike the satellite systems of Eris, Haumea, and Quaoar, for which satellite colours are highly correlated with the colour of the primary \citep{Brown2010Orcus}.  
The colour difference between Orcus and Vanth is much larger than the typical color differences observed in smaller KBO binaries, which also points toward a possible impact origin \citep{Benecchi2009}.

\subsection{Higher Density, Small Moons}
Because Eris, Quaoar, and Haumea have similar $q$ values, it seems likely that all three bodies suffered a similar type of collision.  After the discovery of Haumea, the first collisional family in the Kuiper belt \citep{BrownFamily2007}, numerical simulations were performed to study its origin \citep{LeinhardtHaumea}.  To date, no numerical simulations have been performed for the origin of the Eris and Quaoar systems.  

\citet{LeinhardtHaumea} find that scenarios yielding the proper dynamical state for the system involve the collision of two differentiated ice/rock bodies $R_i=R_t=$650 km and $M_i=M_t=2.25 \times 10^{24}$ grams, a little bit larger than Orcus.  
The collision occurs at $v_{imp}\sim 0.8$ to 0.9 km/s, comparable to the Pluto/Charon impact, with little melting or vaporisation.  
In the half-dozen cases that successfully reproduce a rapidly spinning primary with small icy moons, the impactor and target actually collide twice \citep{LeinhardtHaumea}.
The first collision gravitationally binds the impactor to the target and sets the system spinning; the second collision is a merger.  
When the impactor and target merge, the rock cores of both objects merge.  
This behaviour has also been observed in the impact that formed Earth's Moon \citep{Canup2004}.
The central body spins more rapidly after the rock cores merge, due to the decrease of its moment of inertia.  
A bar instability, similar to that observed in galaxy mergers (e.g., \citealt{BarnesGalaxies2002}) develops, which lofts mantle ice into orbit.  
The orbiting ice clumps together to make Haumea's small moons.
Statistically, such collisions are expected to be common in the scattered disc \citep{LevisonHaumea2008}.  

Unfortunately, this scenario does not reproduce the rock-rich composition of the Haumea primary, nor is it a particularly realistic collisional scenario.  
The successful cases begin with impactor and target densities of 2.0 g/cm$^2$ (which we now know is actually more rock-rich than either Pluto or Charon), and yields primary densities only marginally increased to 2.1 and 2.2 g/cm$^3$, much lower than the 2.6 g/cm$^3$ observed \citep{LeinhardtHaumea}.
Secondly, the only scenarios that yield the rapidly spinning primary and the proper moon masses are those between \emph{identical} objects ($\gamma=0.5$), with equal masses and radii.  
This is highly unlikely.
Moreover, an idealised numerical impact simulation, a perfectly symmetric initial condition such as that used in the Haumea impact can only yield a symmetric outcome: a disc \citep{RobinPers}.  

With new constraints on the sizes and compositions of Eris and Quaoar, it may now be worthwhile to search for impact conditions that could have modified the densities and formed the satellites of these systems.  Before searching for candidate impact scenarios, it is important to determine why the Haumea simulations were so unsuccessful.  Why don't numerical impact simulations yield rock-rich primaries?  Intuitively, high-velocity grazing impacts could plausibly ``chip off'' fragments of ice from the mantle of a fully differentiated primary body, leaving behind a rock-enriched primary with a rapid spin \citep{BrownFamily2007, Brown2002UX25}.  Indeed, if such a collision occurred in the Kuiper belt, any fragments launched from the primary at $v_{esc}$ would soon become physically and dynamically separated from the primary because the Keplerian orbital velocities are similar to $v_{esc}$ for many dwarf planet systems.  This has been demonstrated for the Haumea family by \citet{Lykawka2012}.  

Despite efforts to demonstrate this numerically, the mode of ice loss in dwarf planet collisions remains unclear.  We suggest two reasons why this might be the case.  One possibility is that we have not yet identified the proper impact conditions (velocity, $\gamma$, impact angle, differentiation state of precursors).  Because numerical simulations have explored the origin of only two dwarf planet systems, it seems likely that there are significant areas of parameter space that remain unexplored.  For example, an improbable, but not impossible, head-on collision between two 840 km objects with $\bar{\rho}=1.5$ g/cm$^3$ at $v_{imp}\sim 5v_{esc}$ can create a primary with $\bar{\rho}=2.2$ g/cm$^3$ \citep{Barr2010TNO}.  Thus, exploring a wider variety of impact conditions could be a fruitful avenue for future research.

Another possibility is that the material strength of ice, which has been ignored in all prior simulations of dwarf planet collisions, could modify how the bodies respond to compression during the collision.  Strength effects are thought to be important if the yield stress of the material $\sigma_Y \sim \varrho g R$, where $g$ is the local gravity \citep{Melosh1989}, a condition not met for rock or ice on the dwarf planets.  However, material strength is known to change the partitioning of compressional deformation and heat in the early stages of the impact \citep{Grady1980}, and can lead to radically different outcomes in laboratory \citep{StickleSchultz2011} and numerical \citep{SchultzCrawford} collisions between spherical objects.  The effects could be pronounced in a differentiated body composed of rock and ice, for which $\sigma_Y$ varies by an order of magnitude.  Recent efforts to compare outcomes of icy body collisions with and without strength show that while the compositions of the final bodies may be somewhat similar in both cases, there is substantially more ice fragmentation in the cases including strength \citep{Maindl2014}.  Notably, \citet{Maindl2014} report a single unusual Haumea-like outcome in which a differentiated ice-rock body is spun beyond its breakup limit, permitting its ice mantle to fracture and loft into orbit, losing 15\% of the total system water content in the process.

\section{Primordial Density of Kuiper Belt Material}
%
If Pluto and Charon represent samples of the source material for the Kuiper Belt objects, we can estimate an overall bulk density for that material, $\varrho$, using an average density for the system, $\rho_{sys} = (M_P +M_C)/(V_P + V_C)$, where $M_P$ and $M_C$ are the masses of Pluto and Charon, and $V_P$ and $V_C$ are their volumes.  
Updated values from \emph{New Horizons} gives $\varrho=1.842$ g/cm$^3$.  Alternatively, one could also view the individual densities of Pluto and Charon as an upper and lower bound on the primordial density, yielding $\varrho = 1.781 \pm 0.08$ g/cm$^3$.  In either case, it seems $\varrho \approx 1.8$ g/cm$^3$.  This is not dissimilar from the bulk system densities of the Uranian satellites, $\rho_{sys}=1.63$ g/cm$^3$, which are similar in size to the dwarf planets.  We note there may be a continuum of KBO densities up to our proposed primordial density for the dwarf planet-sized bodies. Mid-sized bodies like Varda and Salacia have densities greater than 1  g/cm$^3$ and less than 1.5 g/cm$^3$ \citep{StansberryIcarus2012, GrundyIcarus2015}. Bodies with radii less than $\sim$ 400 km, including 2002 UX25 and Typhon, have measured densities less than 1 g/cm$^3$ \citep{, Grundy2008, Brown2002UX25}. 

This range of bulk densities implies a rock mass fraction $\sim 72$\% for a nominal ice density $\rho_i=0.92$ g/cm$^3$ and rock density $\rho_r=3.0$ g/cm$^3$.  We have chosen a rock density mid-way between $\rho_r=3.3$ g/cm$^3$ for Prinn-Fegley rock, and $\rho_r=2.8$ g/cm$^3$, the grain density of CI chondrite, thought to be the rocky component of the icy satellites \citep{MuellerMcKinnon}.  The rock density may be lower if the rock is hydrated, but this is unlikely because both bodies would get hot enough to dehydrate silicate in the first billion years of their evolution \citep{Prialnik2015}.  Compression of rock and ices at depth is not a significant effect in bodies of this size \citep{LupoLewis1979, Brown2002UX25}, but the ice density may also be higher if the ice contains significant hydrocarbons \citep{McKinnon_KBO_Book}, which could be tested with \emph{New Horizons} surface spectra.

How can one create the high-density dwarf planets from this primordial material?  Eris can achieve its present density  of 2.52 g/cm$^3$ with a single core-merging collision between two primordial-composition objects with combined masses $M_T=2.1 \times 10^{25}$ g, each of which would be roughly the size of Pluto.  
If only 15\% of the total mass was lost, and that mass was composed of pure ice, Eris could achieve its present-day mean density.  
Quaoar's present density $\rho=1.99$ g/cm$^3$ can be achieved in a core-merging collision between two $\varrho$ primordial objects of radii 405 km, close to the estimated size of Vanth, and smaller than Charon. 
Similarly, loss of 10\% of the total impacting mass, in ice, would give Quaoar the proper density.  
The Haumea-forming collision similar to that suggested by \citet{LeinhardtHaumea} could yield a primary of proper density if $\sim 20$\% of icy material was lost from the system.  
It is interesting to note that these loss rates are not dissimilar from those observed in by \citet{Maindl2014}, and that the amount of ice removed from Haumea and Quaoar impacts is comparable in magnitude to the mass of Salacia.


\section{Triton}
\label{sec:Triton}
Triton's inclined and retrograde orbit strongly suggests that it is a captured satellite \citep{McKinnon1984, GoldreichNeptune, AgnorPluto,Nogueira2011}. The current favoured scenario for capture involves an encounter of a binary Kuiper Belt system with Neptune, in which Triton represents one member of the binary; the other having been lost \citep{AgnorPluto}. Early works on the capture scenario suggested that if Triton were a captured Kuiper Belt object, its ice/rock ratio should be similar to Pluto and thus similar to the primordial densities of the other distant dwarf planets \citep{McKinnonMueller1989}. This viewpoint was bolstered by the remarkable equality between the density of Triton, $2.06$ g/cm$^3$, and the best pre-encounter estimates of Pluto's density from multi-body orbital fits, $2.06$ g/cm$^3$ \citep{Tholen2008}. 

For most of the age of the Solar System, Triton has been in a different environment than the dwarf planets.  Thus, direct comparisons with the dwarf planets may not be valid.  With the updated density for Pluto, it seems that Triton contains more rock than Pluto and has a density closer to Quaoar.   Here, we {\Amy present a simple argument to show that Triton could have lost volatiles during an epoch of intense tidal heating} during its orbital evolution post-capture,  {\Amy which could have modified its ice/rock ratio.}

After capture, Triton would have experienced an episode of strong tidal heating \citep{McKinnon1984, GoldreichNeptune, RossSchubert1990, Correia2009, Nogueira2011} as its orbit around Neptune decreased from an original semi-major axis of $a_o\sim 80$-1000 $R_N$ to its present day value of $a_f=14.3 R_N$ \citep{Nogueira2011}. The vast majority of this energy is dissipated within Triton \citep{McKinnon1984}. The total energy per unit mass dissipated in Triton to date is equal to the difference between the specific orbital energy of the captured object and Triton's present orbit \citep{McKinnonNeptuneBook},
\begin{equation}
\Delta E = \bigg(\frac{GM_N}{2a_f} - \frac{GM_{N}}{2a_o}\bigg),
\end{equation}
where $G$ is the gravitational constant, $M_{N}$ is the mass of Neptune, $a_o$ is Triton's semi-major axis from Neptune at capture, and $a_f$ is the semi-major axis of Triton's orbit around Neptune today. The amount of energy is relatively insensitive to the assumed value of $a_o$ because $a_o>> a_f$.  Regardless of the details of the capture scenario, $\Delta E \approx 10^{11}$ erg/g, a factor of 3 higher than the latent heat of sublimation of water ice $L_s=2.8\times10^{10}$ erg/g \citep{Kieffer2006}.  Alternatively, for a specific heat $C_p\sim 10^7$ erg/g K appropriate for rocky and icy material, this gives a temperature rise $\Delta T \sim 10^4$ K, sufficient to melt Triton's ice and rock \citep{McKinnonNeptuneBook}.  

Numerical integrations of the tidal evolution equations coupled with thermal evolution models show that the vast majority of the tidal heat is dissipated in a short period of time, with strong heating lasting only a few hundred million to a billion years \citep{McKinnon1984, GoldreichNeptune, RossSchubert1990, Prockter2005}.  During this brief epoch of strong heating, the surface heat flow peaks at $F\sim 0.5$ W/m$^2$ (500 erg/s cm$^2$), about half the observed heat flow from Io (e.g.,\citealt{Veeder1994}).  

Heat flows of this magnitude, $\sim 0.25$ W/m$^2$ (250 erg/s cm$^2$) \citep{Howett2011} have also been observed in the south polar terrain of Saturn's icy moon Enceladus, which is experiencing an episode of high tidal heating \citep{MeyerWisdom2007} and concomitant ice loss \citep{Porco2006, Hansen2006, Waite2006}. Much of the thermal emission from Enceladus is coming from four linear features \citep{Spencer2006} about 130 km long and 2 km across \citep{Porco2006}, which are also the source of water-rich plume eruptions, that are removing $\dot{M}_E \sim 2\times10^5$ grams of water-rich material from the satellite per second \citep{Hansen2006}.  

A rough estimate of the efficiency of conversion from tidal heat to sublimation and loss from an icy body can be obtained from the present energy budget of Saturn's moon Enceladus. We construct a crude ``efficiency factor'' using a ratio between the measured mass loss rate times the latent heat and dividing by the tidal heat power, $\dot{E}_{tidal} \sim (\Delta E/\tau)$,
\begin{equation}
f\sim \frac{\dot{M}_E L_s}{\dot{E}_{tidal}}
\end{equation}
which gives $f \sim 0.036$.
If Triton uses $f=3.6$\% of its tidal power to sublimate and lose ice, {\Amy and none of the sublimated ice re-accretes onto Triton}, the total mass lost over a $\tau\sim500$ Myr episode of intense heating is $2.5 \times 10^{24}$ grams, about half of the present-day mass of ice in Triton.  Adding back this mass of ice would yield a body with density $\rho=1.84$ g/cm$^3$, similar to Pluto, and radius $R\sim$ 1455 km, within the range of other dwarf planets.  Although we do not know the exact devolatilization history of Triton, we show that the thermal energy dissipated during its post-capture orbital evolution is capable of modifying a pre-capture Triton from an original  Pluto-like primordial density of $\sim1.8$ g/cm$^3$ to its present composition, depending on the value of $f$.  

A range of binary mass ratios produce a viable capture Triton scenario, but we note that the all modelling attempts have used the present-day mass of Triton for one of the bodies in the binary. The most successful capture scenarios from \cite{DavidV2008} require a pre-capture binary $q$ of $\sim 0.25$ to $0.333$. During Neptune's subsequent planetesimal-driven migration, \cite{Nogueira2011} can produce successful captures with a escaping binary companion with 0.1, 0.3, 1, and 3 Triton masses (implying $q$ values of 0.09, 0.23, 0.5, and 0.25 respectively). This could place the pre-cursor binary system in density-$q$ space close to Pluto/Charon and Orcus/Vanth, the two other systems we propose formed in a gentle collision, potentially enabling Triton to have our proposed primordial density at the time of Neptune capture. 

Regardless of the exact scenario for Triton's origin, we show that the tidal heating from its initial orbital evolution is sufficient to remove a substantial amount of ice from the body.  Thus, Triton's present composition may not be primordial, and should not be used as a standard by which to interpret the densities of the largest bodies in the Kuiper belt.

\section{Conclusions}
We now have good estimates for the sizes and densities of the largest objects in the Kuiper belt, the dwarf planets with absolute magnitude less than 3, corresponding to radii $\gtrsim500$ km.  Many of these objects have moons, which has facilitated estimates of the mass of the central body.  Collisions have undoubtedly played a role in shaping the composition of the distant dwarf planets.  We have shown that these systems fall into two broad categories, and we hypothesise that these categories represent the products of two types of planet-scale collision.  Pluto/Charon and Orcus/Vanth comprise one category, formed from a gentle, grazing collision with little to no vaporisation, melting, or heating, and very little mass loss.  We propose that the composition of these bodies may represent the bulk composition of the primordial material from which the Kuiper belt objects accreted, with density $\varrho\sim 1.8$ g/cm$^3$.  Triton could have been composed of this primordial material at the time of capture, but vigorous tidal heating and volatile loss could have modified Triton's composition.

Eris, Haumea, and Quaoar, with small satellites and high rock fractions, comprise the other category.  Collisions between Orcus-to-Pluto-sized objects, in which a few tens of percent of the system is carried away in icy fragments, could explain the ice-rich compositions of these three bodies.  However, numerical simulations have so far failed to produce such outcomes.  We suggest that the gap between model and observation could be bridged by a broader exploration of impact parameter space, or by adding physical processes hitherto overlooked, such as material strength \citep{StickleSchultz2011, SchultzCrawford, Maindl2014}.  

If significant ice was lost from Eris and Quaoar in collisions, these objects may have had (or presently have) collisional families.  Identification of collisional families in the Kuiper Belt is difficult because fragments lost during a collision have velocities comparable to the Keplerian orbital velocity, and do not remain in physical or dynamical proximity to the parent dwarf planet (e.g., \citealt{Lykawka2012}).  The Haumea family was identified due to the unique spectral signature of its members \citep{BrownFamily2007}.  It is possible this could provide a means by which collisional families could be identified for Eris and Quaoar.

If our hypothesis is true, we predict that dwarf planets without moons should have a mean density comparable to that of the Pluto/Charon system.  Specifically, we predict that the overall system density of the Orcus/Vanth system is comparable to this primordial density.  Improved estimates of the sizes of Orcus and Vanth are crucial to testing our hypothesis.  Stellar occultation measurements could yield the sizes of Orcus, which would help to further constrain its bulk density.  Also, the present estimate for the density of Quaoar, $1.99 \pm 0.46$ g/cm$^3$, is tantalisingly close to our estimate of the primordial density.  An improved measurement of the mass of Quaoar, which is the main driver of uncertainty in the estimate for mean density, could distinguish whether Quaoar is truly similar to Eris and Haumea (low $q$ and high $\bar{\rho}$), or whether it suffered a different type of collision. {\Amy An additional test of this framework will be the characterization of Makemake's  newly discovered satellite \citep{Parker2016}. At this time, there is no secure orbit for the moon and thus no mass measurement for Makemake. Colour measurements of the satellite have not been reported. If the $q$ of the system is similar to that of Pluto/Charon and Orcus/Vanth, then we expect  Makemake will have a density of $\sim$$1.8$  g/cm$^3$; otherwise the primary will be more rock-rich, similar to Eris and Haumea. } Finally, the overwhelming majority of impact simulations have used $\bar{\rho}\sim 2$ g/cm$^3$ material as a starting point,  have overlooked the possible importance of material strength \citep{Maindl2014}, and have explored only a small area of the total impact parameter space.  More impact simulations with more realistic material behaviors, and a variety of impact conditions and compositional initial states are crucial to improving our understanding of role of collisions in modifying dwarf planet densities.

\section*{Acknowledgements}
Author Barr acknowledges support from NASA PG\&G NNX15AN79G.  We thank Craig Agnor for helpful discussions.
This research has made use of NASA's Astrophysics Data System Bibliographic Services.




\bibliographystyle{mnras}
\bibliography{refs} 


\bsp	
\label{lastpage}
\end{document}